\documentclass[proof]{WileyASNA-v1}

\articletype{Article Type}%

\received{}
\revised{}
\accepted{}

\begin{document}

\title{BH Mass, Jet and Accretion Disk Connection: An Analysis of Radio-loud and Radio-quiet Quasars}

\author[1]{Avinanda Chakraborty*}

\author[2]{Anirban Bhattacharjee**}

\authormark{Avinanda Chakraborty \& Anirban Bhattacharjee}

\address[1]{\orgdiv{Department of Physics}, \orgname{Presidency University}, \orgaddress{\state{West Bengal}, \country{India}}}

\address[2]{\orgdiv{Department of Biology, Geology and Physical Sciences}, \orgname{Sul Ross State University}, \orgaddress{\state{Texas}, \country{USA}}}

\corres{*86/1, College Street, Kolkata, West Bengal 700073.  \email{avinanda.rs@presiuniv.ac.in}\\
**East Highway 90 Alpine, TX 79832, USA.  \email{axb14ku@sulross.edu}}

\abstract{Surveys have shown radio-loud (RL) quasars constitute $10 \%$-$15 \%$ of the total quasar population. However, it is unknown if the radio-loud fraction (RL quasars/Total quasars) remains consistent among different parameter spaces. This study shows that radio-loud fraction increases for increasing full width half maximum (FWHM) velocity of the H$\beta$ and MgII broad emission line. Our data has been obtained from \cite{shen} catalogue. To investigate the reason, in this preliminary study we analyse various properties like bolometric luminosity, optical continuum luminosity, black hole (BH) mass and accretion rate of RL quasars and RQ quasars sample which have FWHM greater than 15000km/s (High broad line). From the distributions we can conclude for all the properties in high broad line, RL quasars are having higher values than RQ quasars. We have predicted RL quasars are intrinsically brighter than RQ quasars and also predicted BH mass-jet connection and accretion disk-jet connection from our results but to conclude anything more analysis is needed.}

\keywords{Surveys, quasars, redshift, bolometric luminosity, optical continuum luminosity, black hole mass and accretion rate, accretion disk, jet.}

\maketitle

\section{Introduction}

Quasars are the most luminous active galactic nuclei (AGN) and are powered by accretion onto supermassive black holes (SMBHs) \citep{salp, bell}. Only 10\% of the total quasars are radio-loud (RL) \citep{sandage}. The main difference between both Radio-loud quasars and Radio-quiet (RQ) quasars is presence of powerful radio jets \citep{bridle, mullin}. However, there is evidence of weak radio jets in RQ quasars also \citep{ulvestad, leipski}. The jet production is yet not understood. Accretion disk and black hole (BH) in active galactic nuclei are thought to be closely connected with relativistic radio jets \citep{Blandf, Payne, dunlop, marconi, blandford, fernandes, lin}.\\
By probing the relation between disk inflow and jet outflow it is possible to investigate the jet production mechanisms from observations, because the ejected jet outflow is a part of the material infalling toward the central BH, and the magnetic field is also accumulated by that accreting material in the vicinity of the SMBHs. Observations have shown that radio emission of quasar is connected with the luminous optical emission lines \citep{Ba, Oster}.\\
Studies have shown that radio luminosity and BH mass are correlated \citep{Ari, Mark, Ross, McLu}. These studies have shown that, radio emission can be widely used as an indirect measurement of energy transported from the central engine through the jets. For active jets, \cite{Bland} and \cite{Falck} have shown radio luminosity varies as  $L_{R} = \dot{M^{1.4}}$ where $\dot{M}$ is the accretion rate. 
\begin{figure*}
\begin{center}
\begin{tabular}{c}
        \resizebox{7cm}{!}{\includegraphics{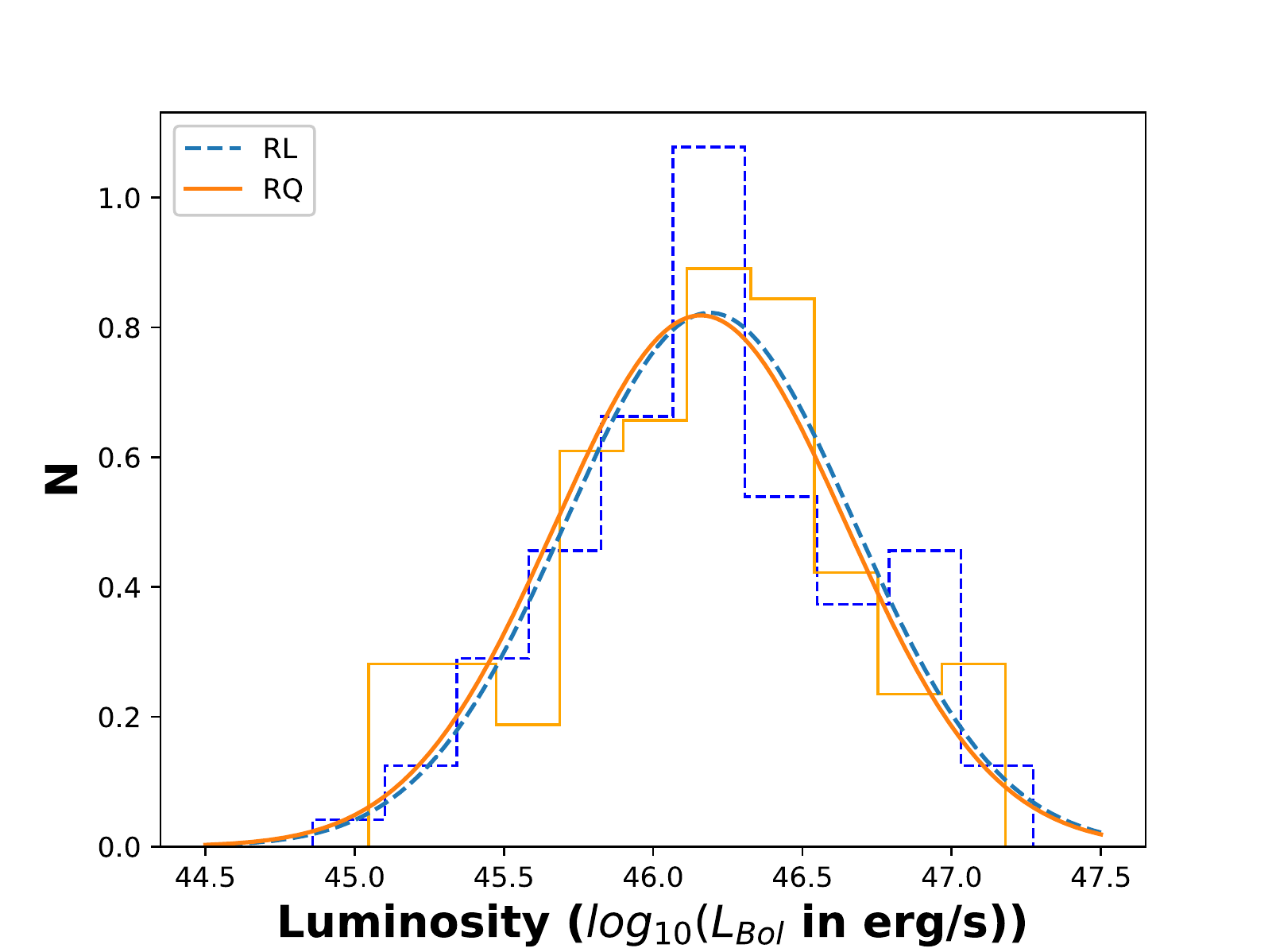}}
        \resizebox{7cm}{!}{\includegraphics{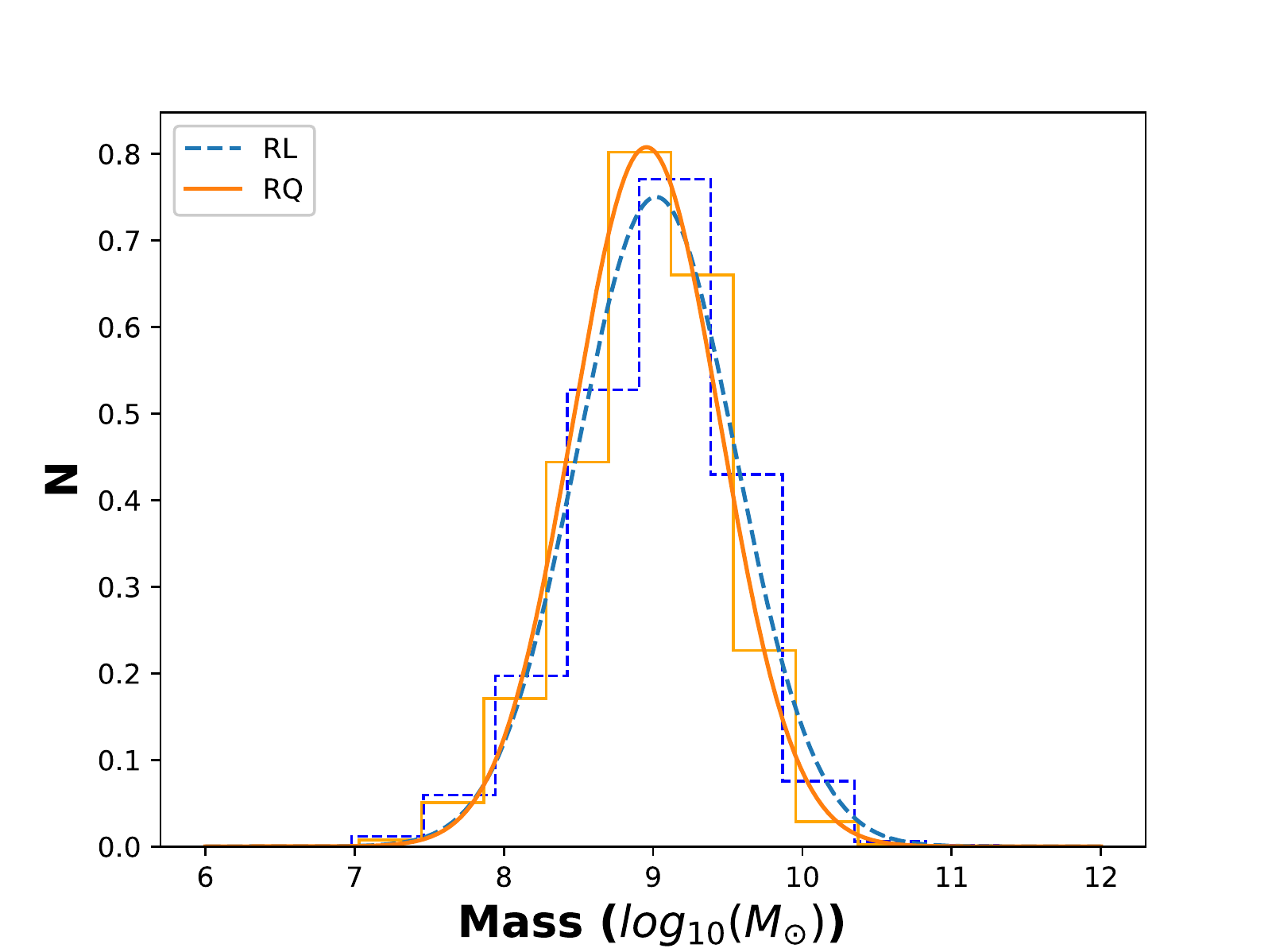}}
\end{tabular}
\caption{{\bf Left} Normalised distributions of bolometric luminosity of all RL quasars and RQ quasars. {\bf Right} Normalised distributions of BH mass of all RL quasars and RQ quasars.}
\end{center}
\end{figure*}

\begin{figure*}
\begin{center}
\begin{tabular}{c}
        \resizebox{7cm}{!}{\includegraphics{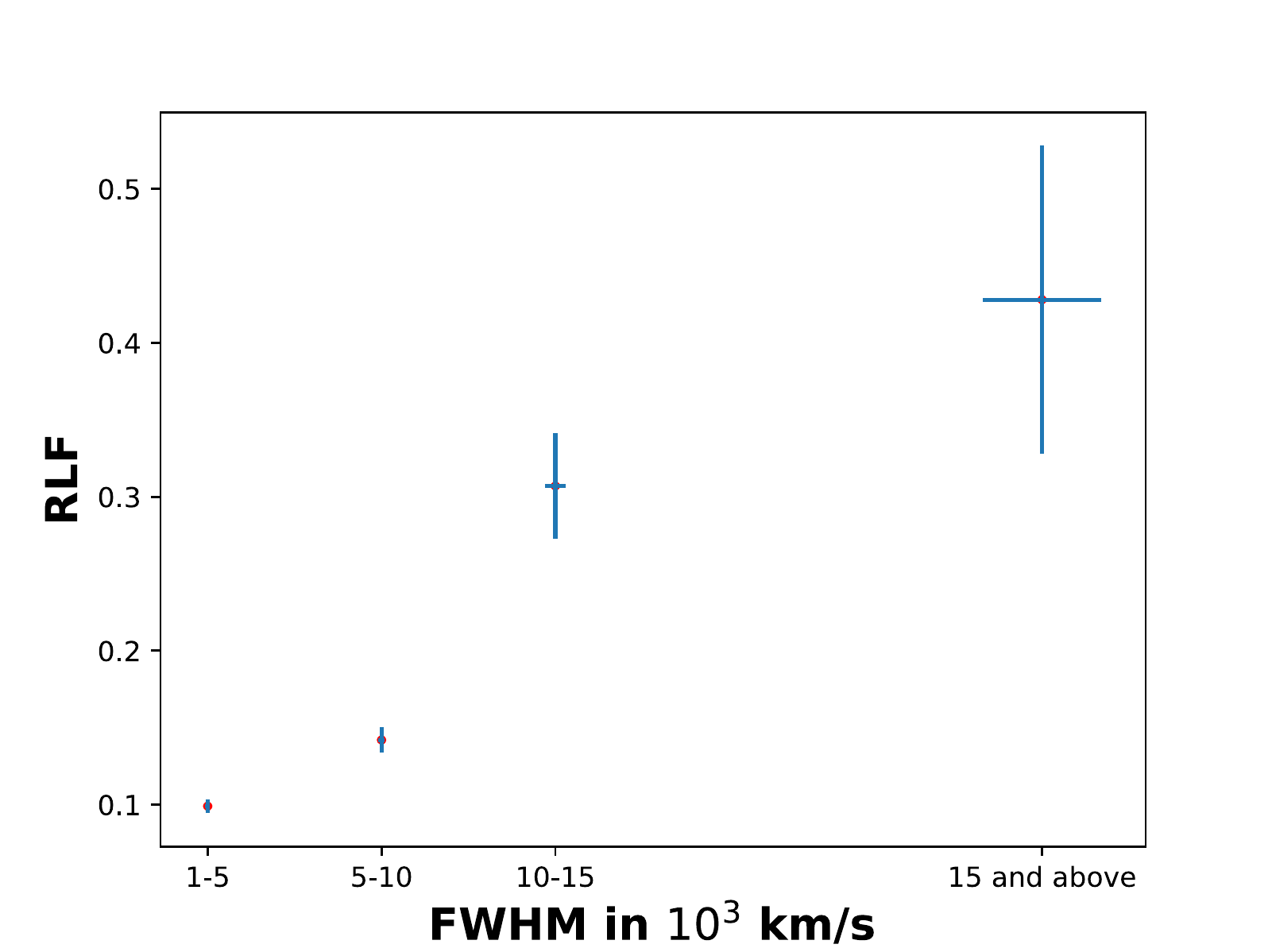}}
        \resizebox{7cm}{!}{\includegraphics{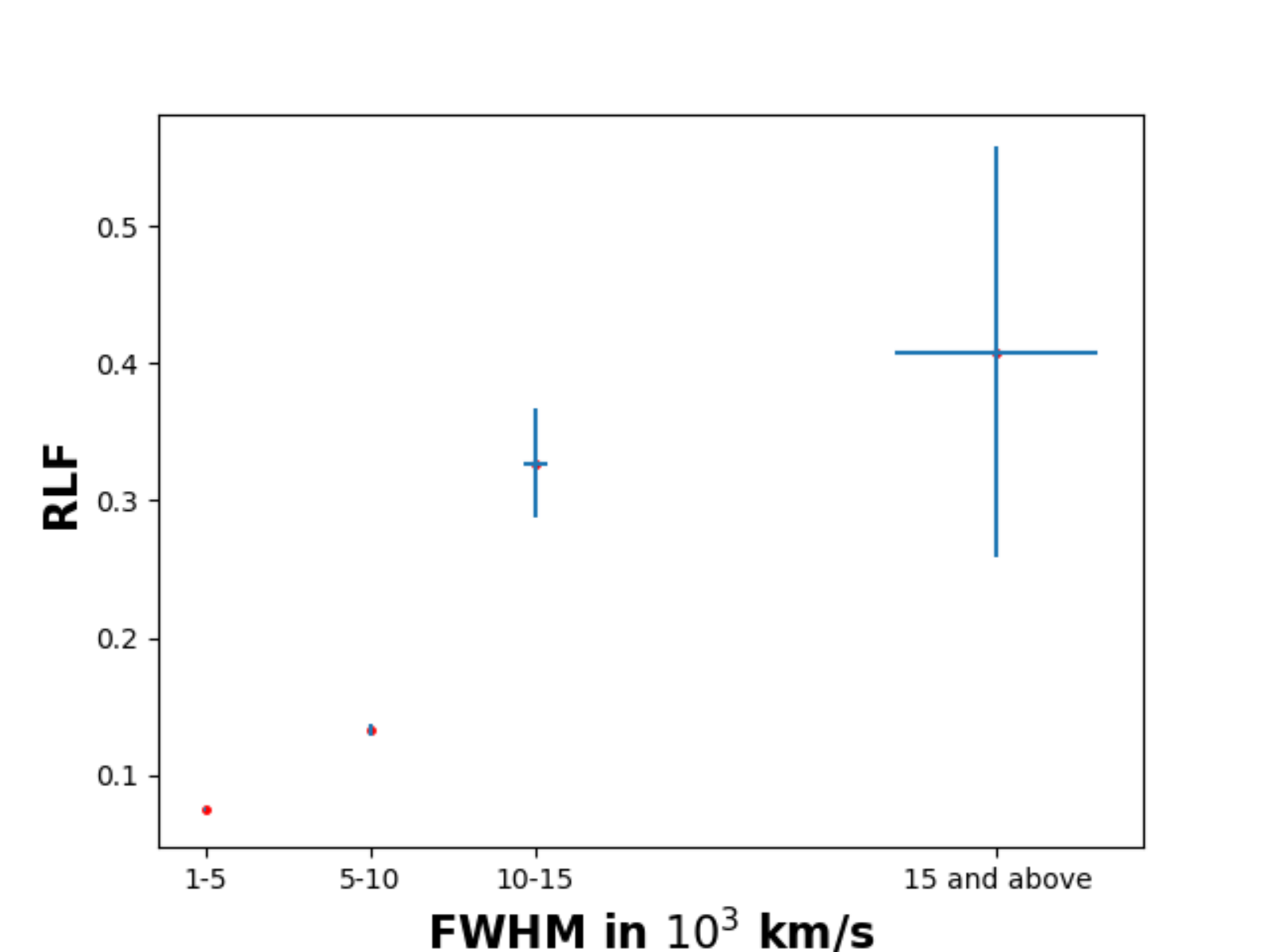}}
\end{tabular}
\caption{{\bf Left} Variation of Radio Loud Fraction across different FWHM of broad H$\beta$. From this plot we can see that the radio-loud fraction increases with FWHM. {\bf Right} Variation of Radio Loud Fraction across different FWHM of broad MgII. From this plot we can also see radio-loud fraction increases with FWHM.}
\end{center}
\end{figure*}

\begin{figure*}
\begin{center}
\begin{tabular}{c}
        \resizebox{7cm}{!}{\includegraphics{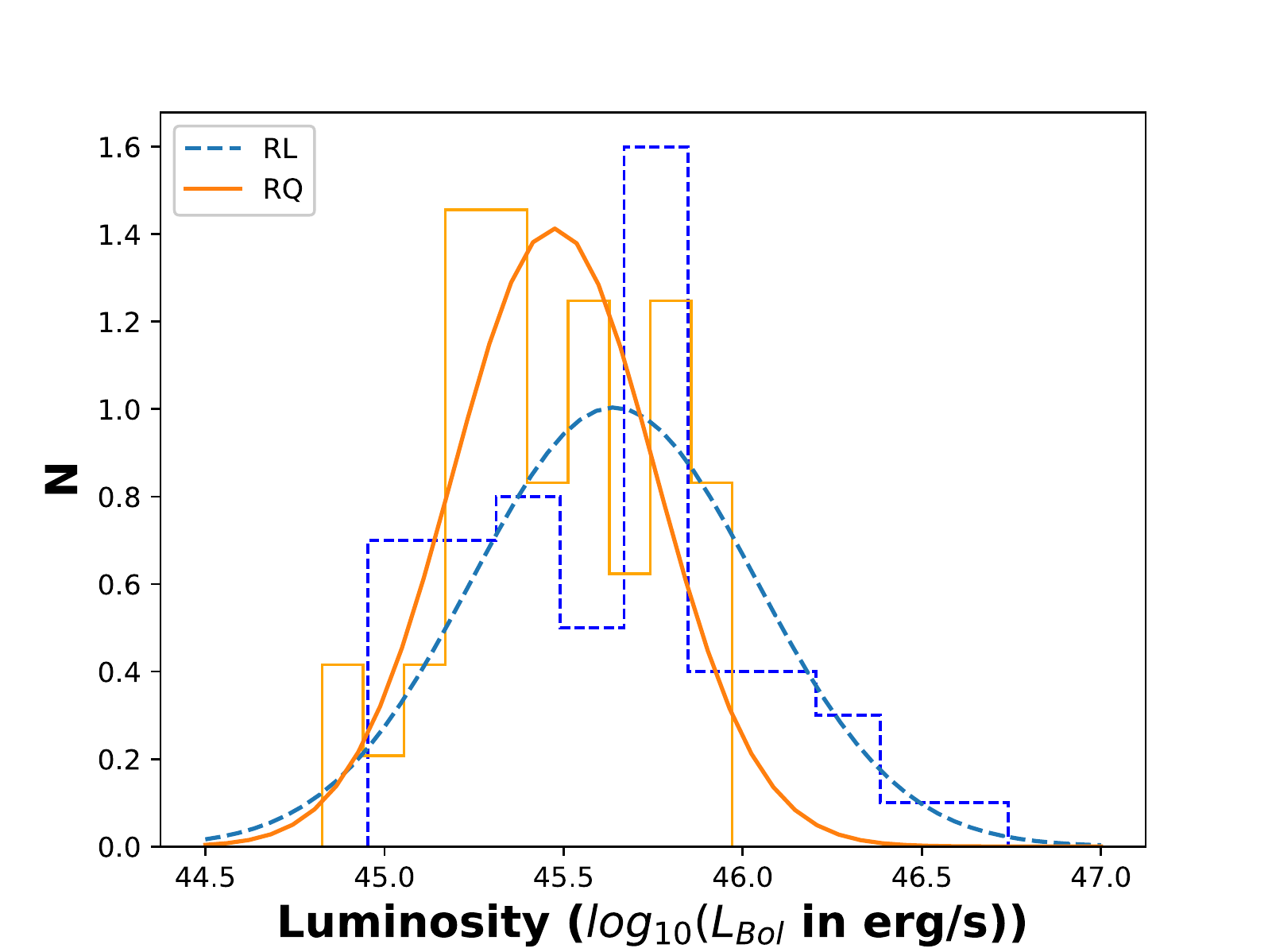}}
        \resizebox{7cm}{!}{\includegraphics{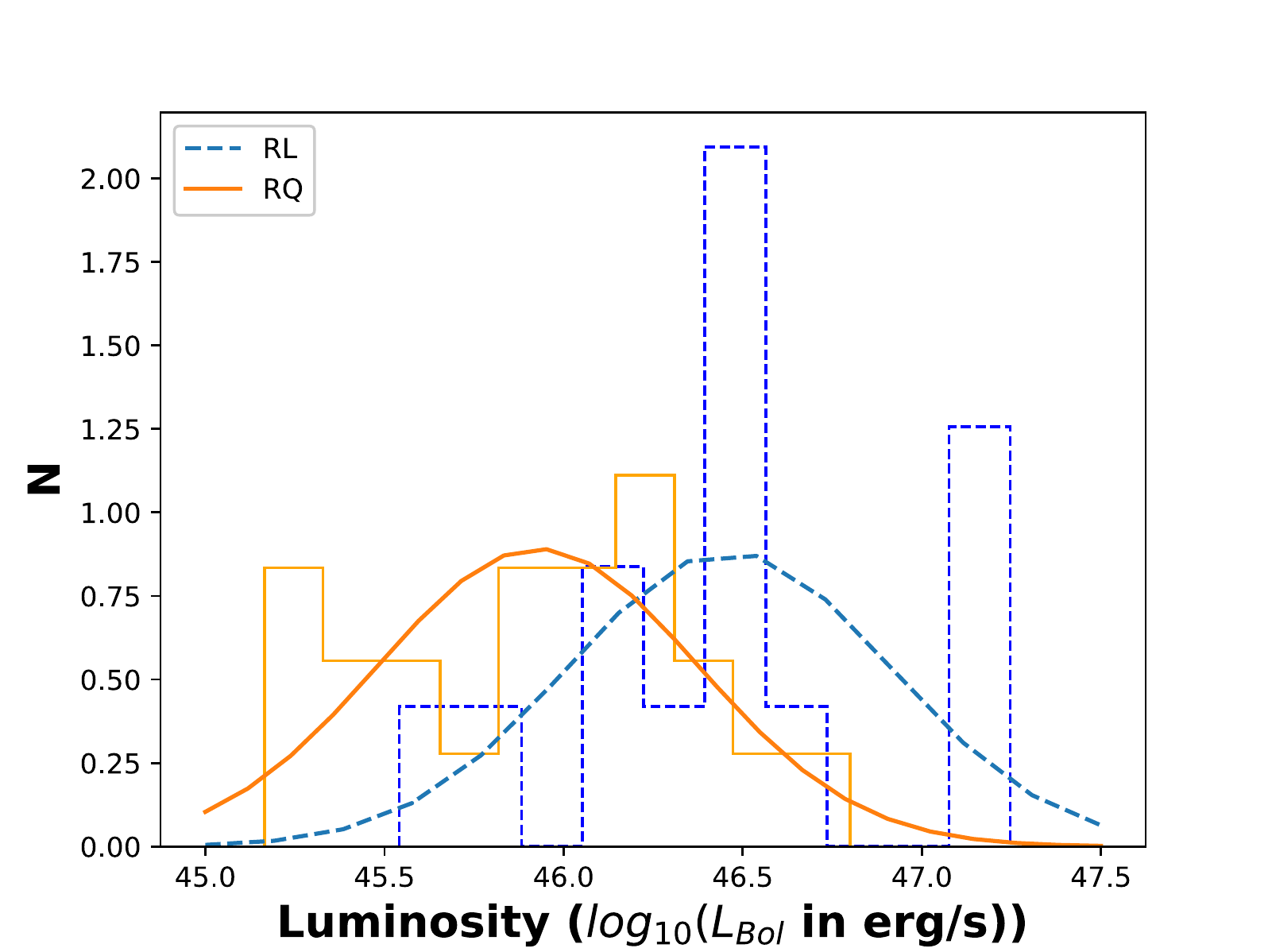}}\\
        \resizebox{7cm}{!}{\includegraphics{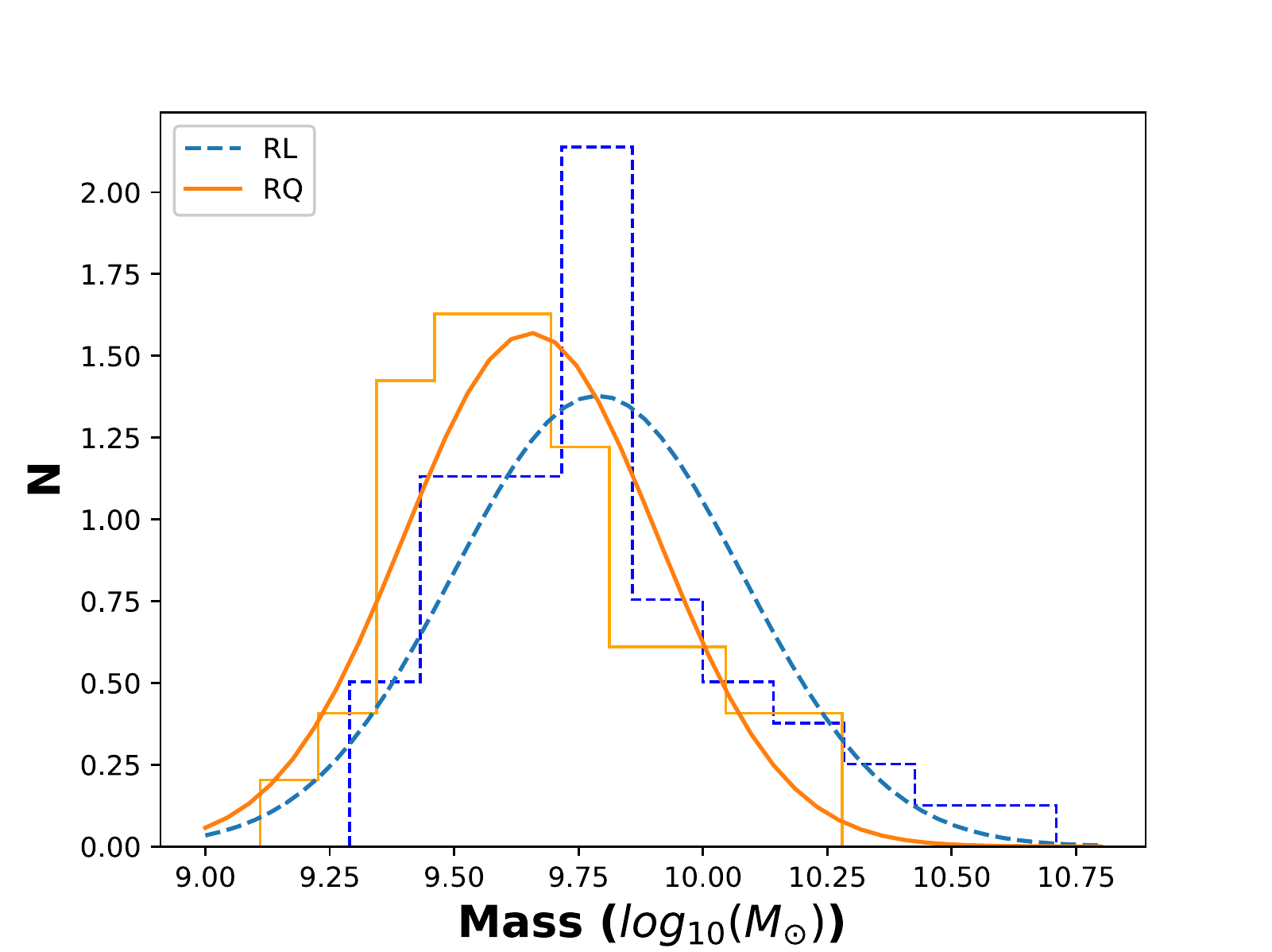}}
        \resizebox{7cm}{!}{\includegraphics{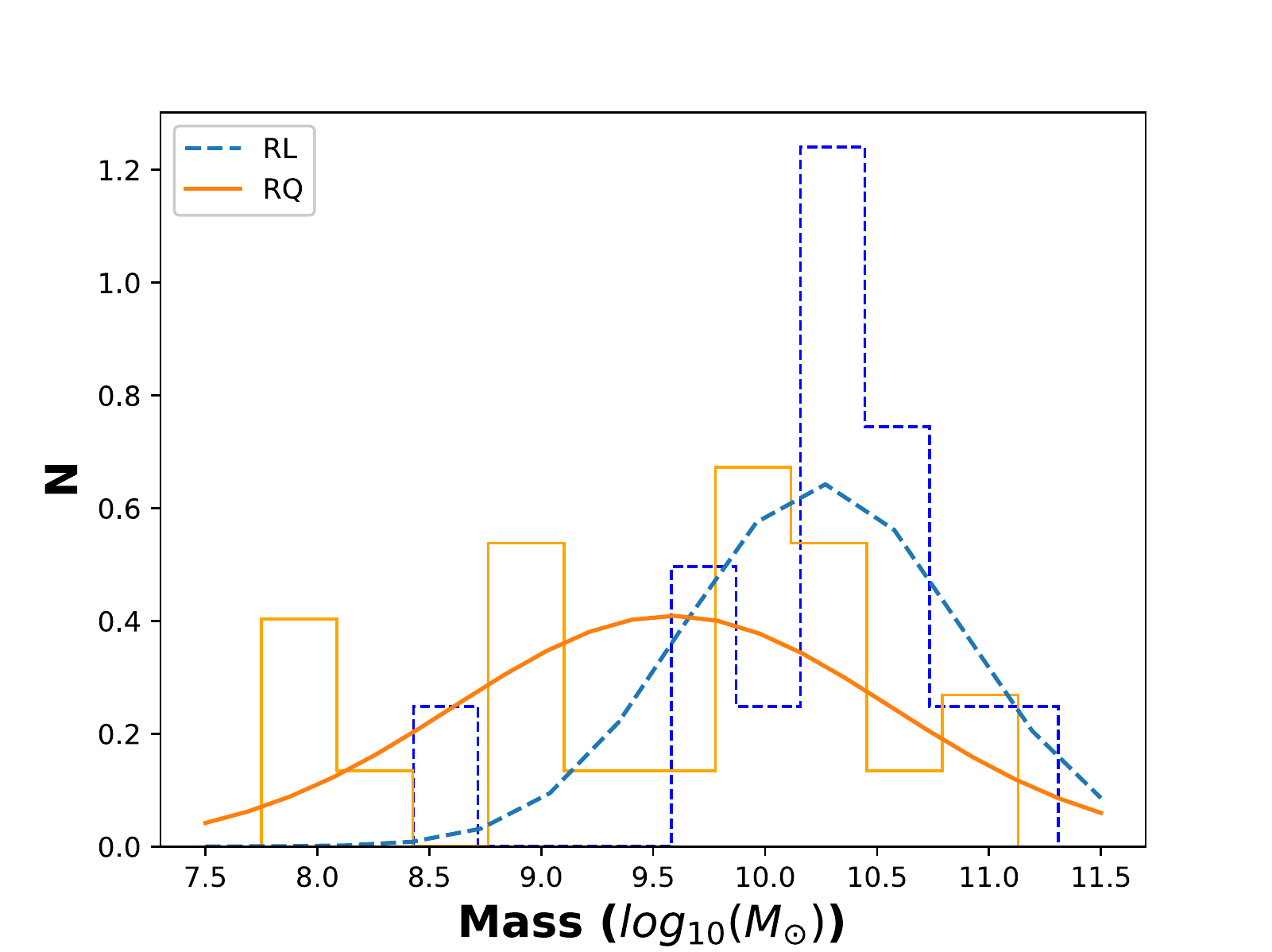}}
        
\end{tabular}
\caption{{\bf Top Left} Normalised distribution of bolometric luminosity of RL quasars and RQ quasars with FWHM greater than 15000 km $s^{-1}$ with Gaussian fits of H$\beta$. {\bf Top Right} Normalised distribution of bolometric luminosity of RL quasars and RQ quasars with FWHM greater than 15000 km $s^{-1}$ with Gaussian fits of MgII.{\bf Bottom Left} Normalised distribution of BH mass of RL quasars and RQ quasars with FWHM greater than 15000 km $s^{-1}$ with Gaussian fits for H$\beta$. {\bf Bottom Right} Normalised distribution of BH mass of RL quasars and RQ quasars with FWHM greater than 15000 km $s^{-1}$ with Gaussian fits for MgII.}
\end{center}
\end{figure*}

\begin{figure*}
\begin{center}
\begin{tabular}{c}
        \resizebox{7cm}{!}{\includegraphics{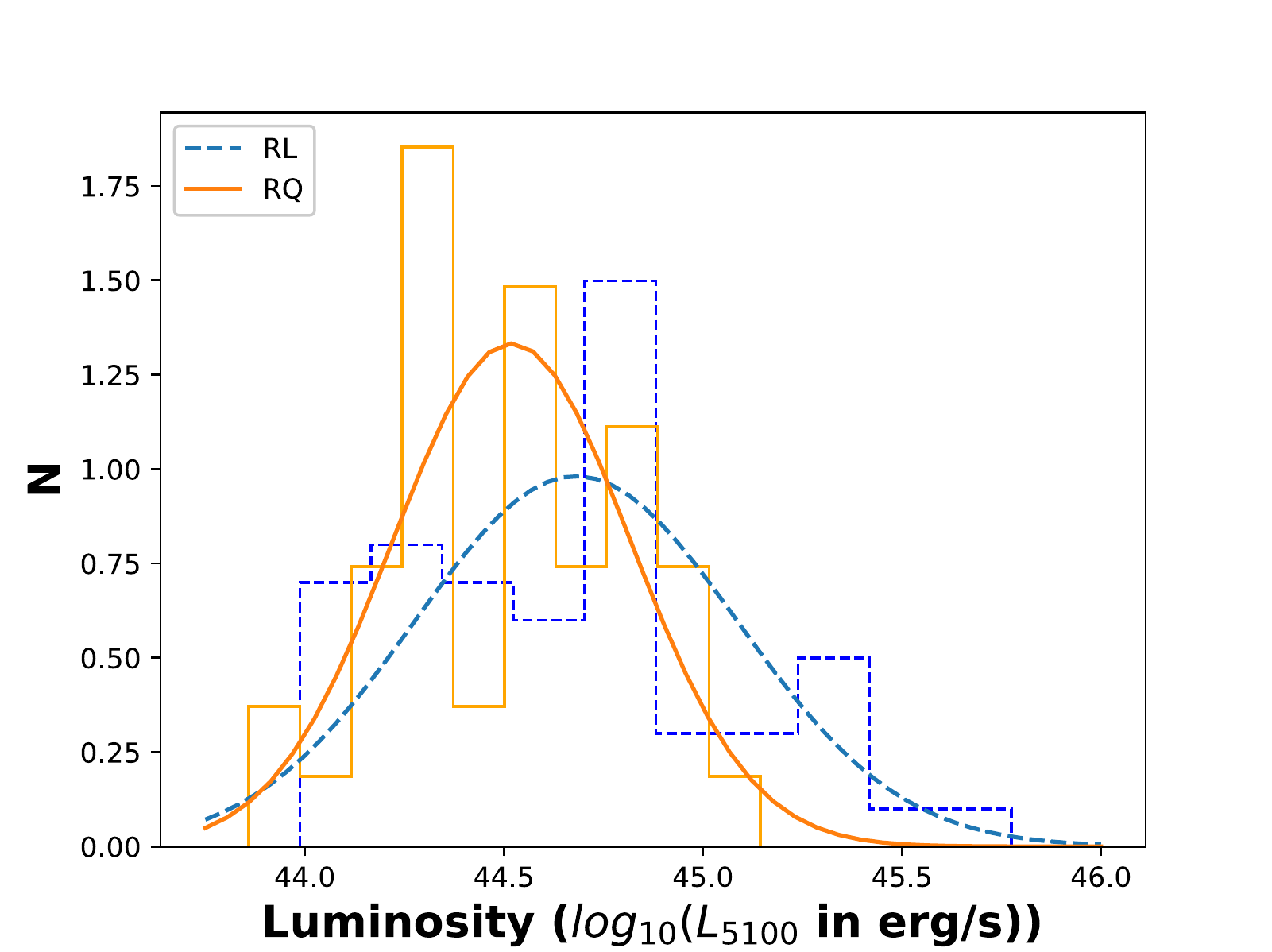}}
        \resizebox{7cm}{!}{\includegraphics{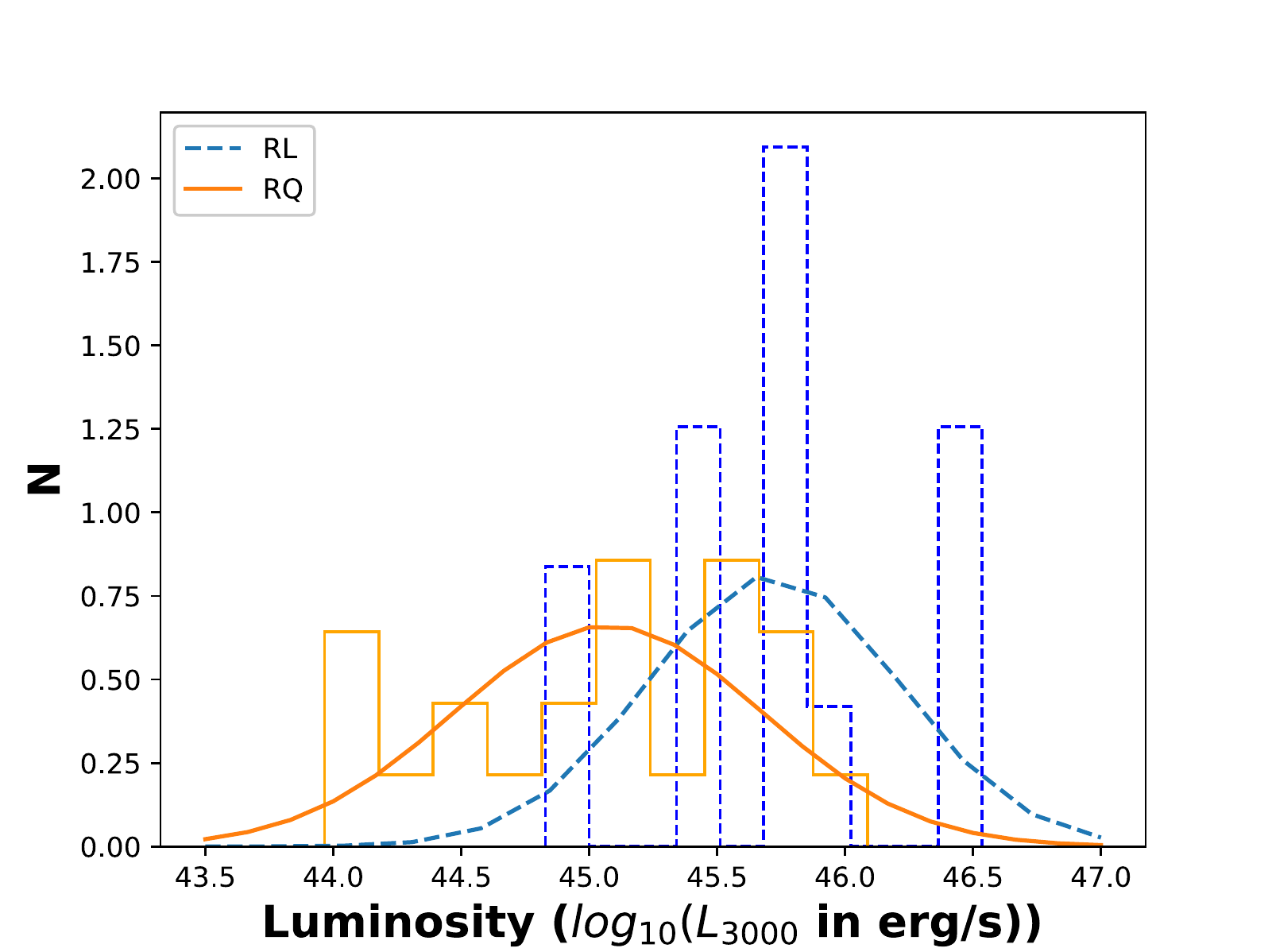}}\\
        \resizebox{7cm}{!}{\includegraphics{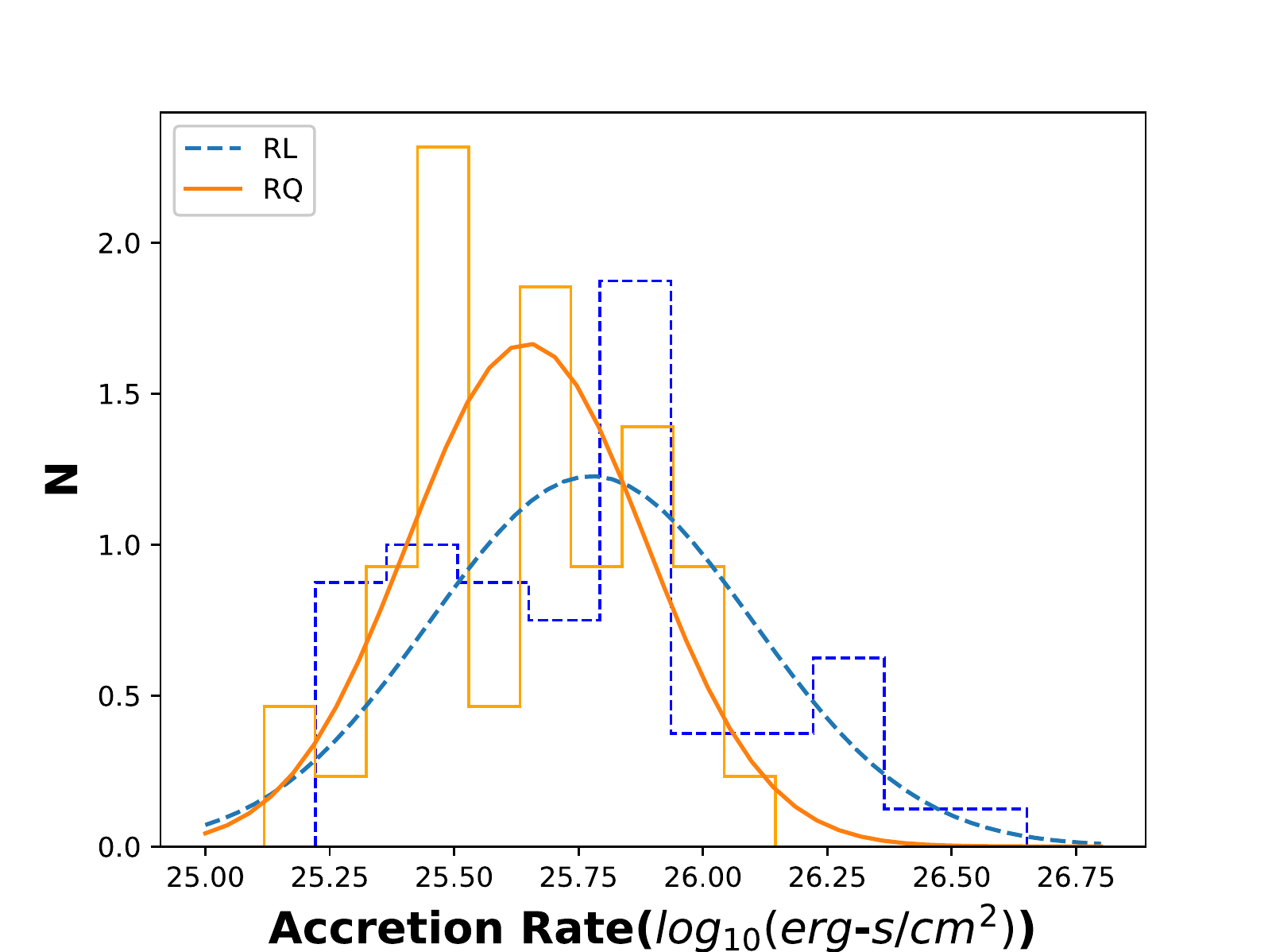}}
        \resizebox{7cm}{!}{\includegraphics{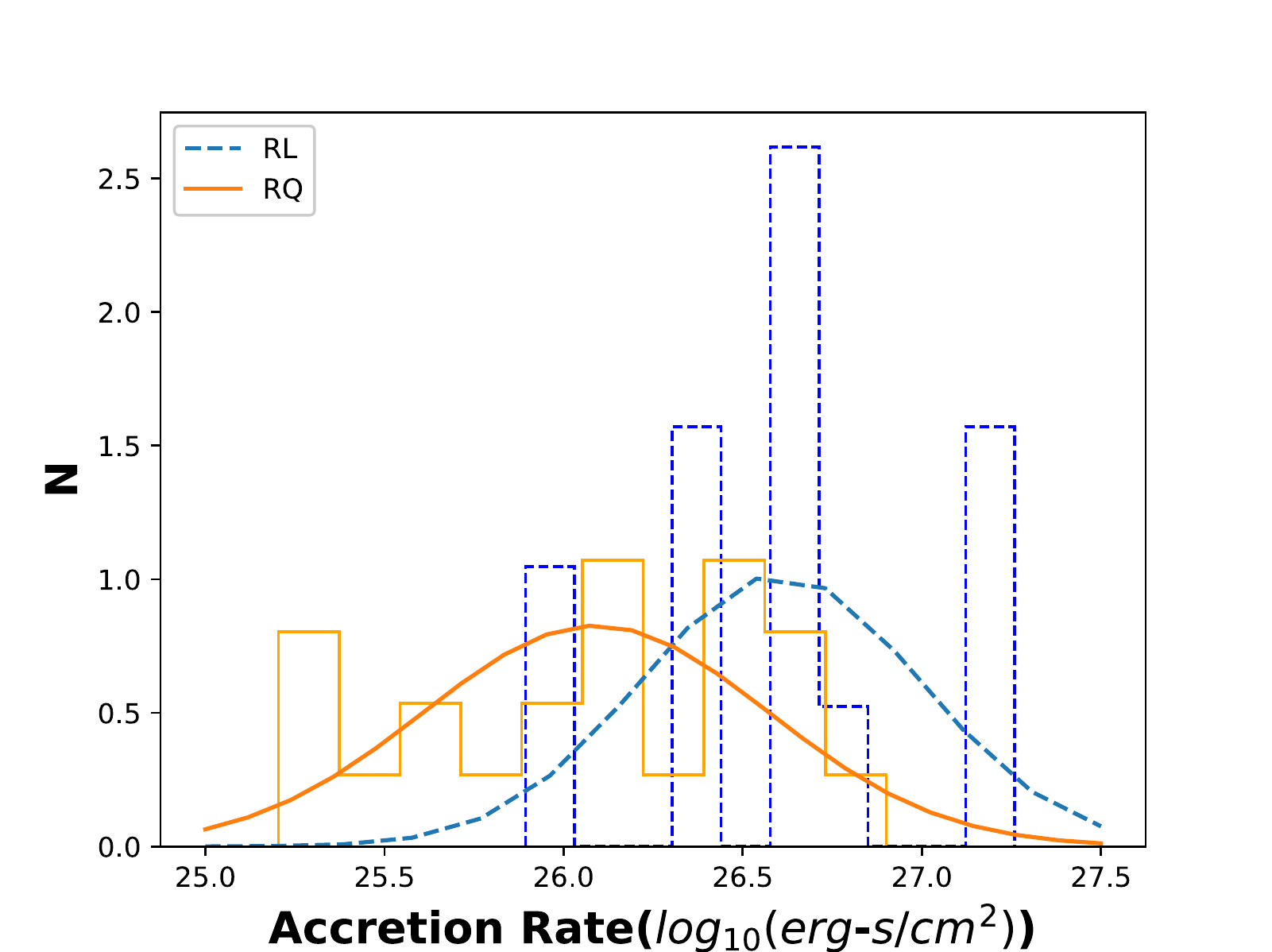}}\\
        \resizebox{7cm}{!}{\includegraphics{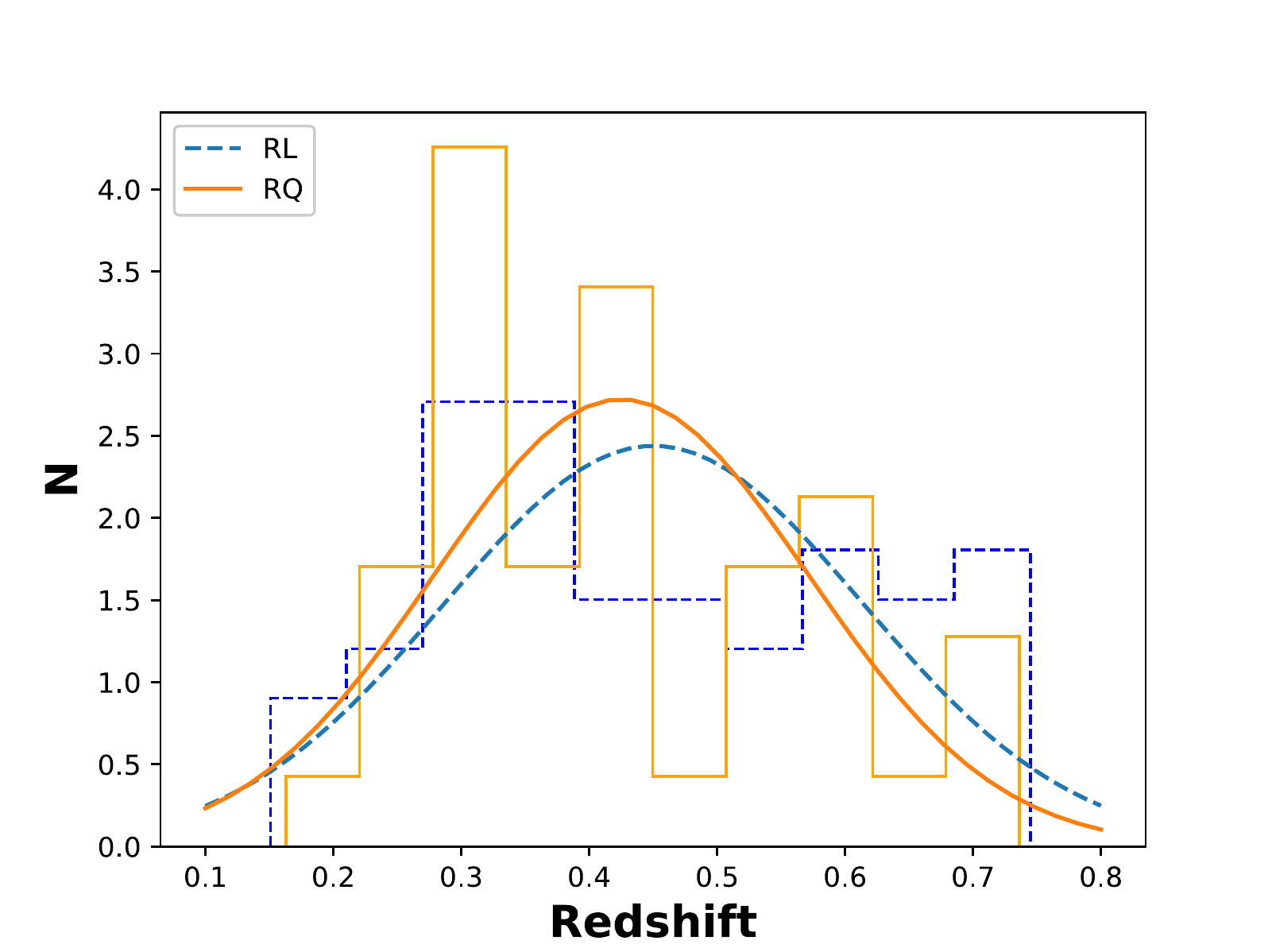}}
        \resizebox{7cm}{!}{\includegraphics{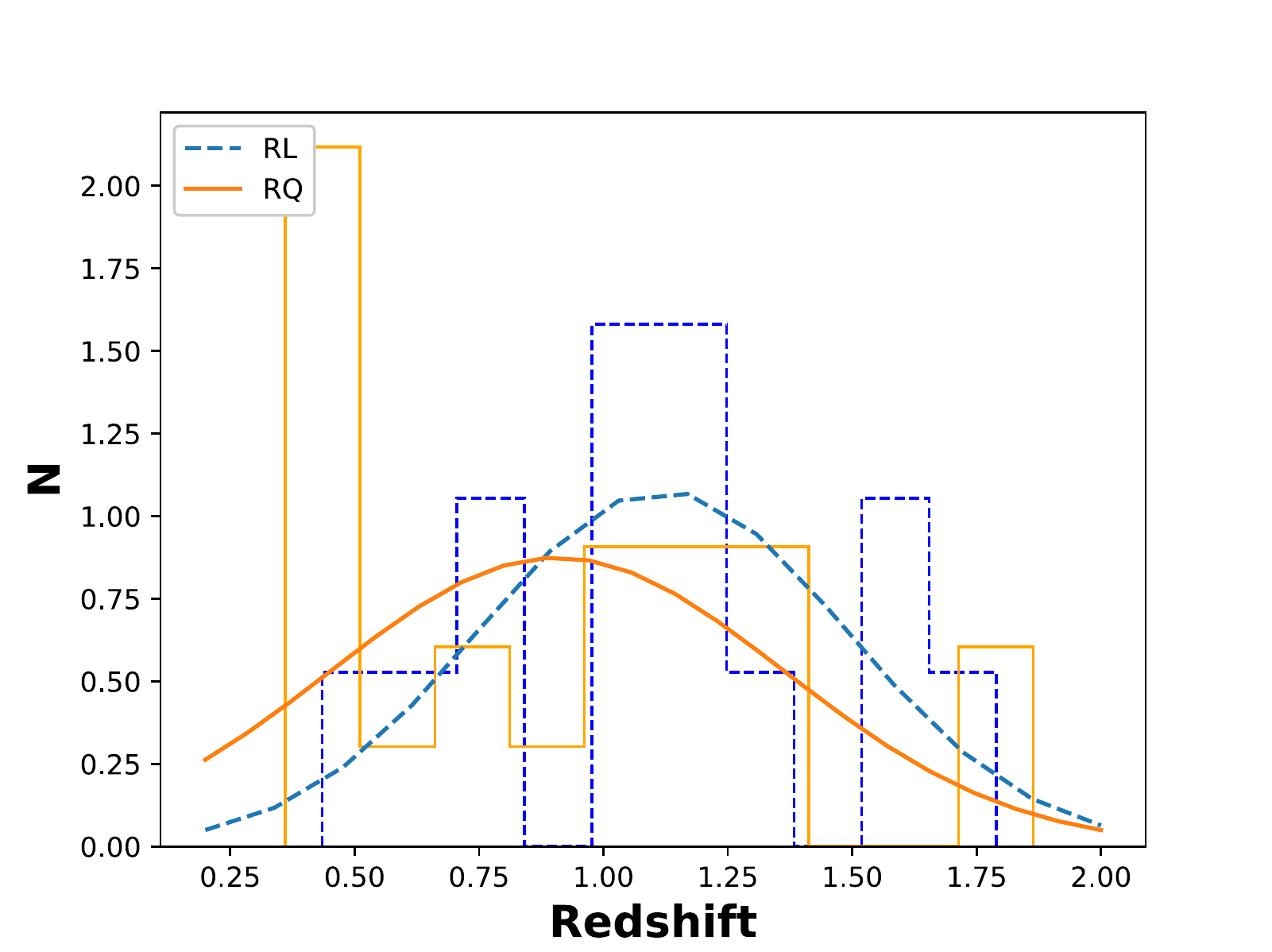}}
\end{tabular}
\caption{{\bf Top Left} Normalised distribution of optical continuum (5100 \.{A}) luminosity of RL quasars and RQ quasars with FWHM greater than 15000 km $s^{-1}$ with Gaussian fits for H$\beta$. {\bf Top Right} Normalised distribution of optical continuum (3000 \.{A}) luminosity of RL quasars and RQ quasars with FWHM greater than 15000 km $s^{-1}$ with Gaussian fits for MgII. {\bf Middle Left} Normalised distribution of accretion rate of RL quasars and RQ quasars with FWHM greater than 15000 km$s^{-1}$ with Gaussian fits for H$\beta$. {\bf Middle Right} Normalised distribution of accretion rate of RL quasars and RQ quasars with FWHM greater than 15000 km $s^{-1}$ with Gaussian fits for MgII. {\bf Bottom Left} Normalised distribution of redshift of RL quasars and RQ quasars with FWHM greater than 15000 km $s^{-1}$ with Gaussian fits for H$\beta$. {\bf Bottom Right} Normalised distribution of redshift of RL quasars and RQ quasars with FWHM greater than 15000 km $s^{-1}$ with Gaussian fits for MgII.}
\end{center}
\end{figure*}
We have visually inspected our quasar spectra to avoid spectra with missing data. So our final sample contains 298 RL quasars and 1,910 RQ quasars for H$\beta$ and 293 RL quasars and 3,244 RQ quasars for MgII. Among RL quasars, 56 and 14 are having high broad line (FWHM $>$ 15000 km$s^{-1}$) and in RQ quasars, 41 and 22 are having high broad line sources in H$\beta$ and MgII respectively. 

In our work, we have investigated if the full width half maximum (FWHM) velocity for broad H$\beta$ and MgII has any influence on radio properties. Our primary goal is to explore why we are getting more number of RL quasars for higher FWHM values. So we have selected one subsample (FWHM $>$ 15000 km$s^{-1}$) of both the lines to investigate different properties like bolometric luminosity, BH mass, optical continuum luminosity and accretion rate of RL quasars and RQ quasars in this sample space and have found RL quasars are intrinsically brighter than RQ quasars. Also we have found a correlation between BH mass and jet and a correlation between accretion disk and jet. Throughout this paper we have used $\Omega_{\Lambda}$ = 0.7, $\Omega_{0}$ = 0.3 and h = 0.7 as the cosmological parameters.

\section{The Sample} 
\subsection{Sloan Digital Sky Survey (SDSS)}
Our main quasar catalogue comes from the SDSS \citep{york} Data Release 7 \citep{abaz} Quasar catalogue \citep{shen}. It consists of 105,783 quasars brighter than Mi = -22.0 and are spectroscopically confirmed. 
We have chosen to use DR7 because it includes line fits with multiple Gaussian and contains all the BH masses for the broad H$\beta$ and broad MgII. Line luminosities, equivalent widths and iron emission strengths etc quantities are also compiled in the catalogue. H$\alpha$ properties are also added in this compilation. Different fiducial scaling relations to compute H$\beta$- and MgII-based virial BH masses are used and except for CIV, different procedures to measure the broad H$\beta$ and MgII have been adopted.  
\\\cite{shen} cross matched the quasar catalogue of SDSS DR7 and FIRST survey. The SDSS contains 105,783 quasars. FIRST matched number is 99,182. Among these 9,399 are RL quasars and 88,979 RQ quasars. We applied redshift cuts of z $<$ 0.75 for H$\beta$ and z $<$ 1.9 for MgII sample. To avoid biasing in the radio properties, the SDSS quasar sample is also restricted to i-band magnitude (i $<$ 19.1).

\section{Method and Results}
Figure 1 shows that the bolometric luminosity distributions and the BH mass distributions of all the RL quasars and RQ quasars of \cite{shen} which tells that there is no such difference in the distributions of RL quasars and RQ quasars.

\subsection{Radio loud fration}
Figure 2 shows the variation of radio-loud fraction across FWHM for both the emission lines.
From figure 2 we can see radio-loud fraction is increasing with FWHM for both the lines which implies, in high broad line region RL quasars are more in number. To investigate the reason now we will analyse different properties of RL quasars and RQ quasars with extremely high FWHM $(> 15000 \ km \ s^{-1})$ for both H$\beta$ and MgII.
\subsection{Analysis of H$\beta$ and MgII line properties}
We have analysed the fundamental properties of RL quasars and RQ quasars for high broad line region for both H$\beta$ and MgII. Figure 3 representing normalised distributions of bolometric luminosity and BH mass of RL quasars and RQ quasars with Gaussian distributions. We can observe that RL quasars have a slightly higher mass and luminosity when compared to RQ quasars.
\\We have also checked optical continuum luminosity for H$\beta$ (at 5100\.{A}) and MgII (at 3000\.{A}) at rest frame and accretion rate for our H$\beta$ and MgII sample.
For our accretion rate calculation we first corrected the bolometric luminosity by calculating bolometric correction \citep{Netz} for each quasar of our sample and then calculated accretion rates using the above mentioned formula. From Figure 4 it can be said that optical continuum luminosity at 5100\.{A} and  3000\.{A} and accretion rate plots are also showing similar features as bolometric luminosity and BH mass plots.
\\Finally we checked the redshift distributions of RL quasars and RQ quasars in high broad line for both H$\beta$ and MgII to check whether there is any selection bias in the sample or not. Figure 4 also represents normalised distributions of redshift of RL quasars and RQ quasars with Gaussian fits for both the lines. For H$\beta$ sample the redshift distributions of RL quasars and RQ quasars are almost identical and for our MgII sample for higher redshift values we are getting more RL quasars than RQ quasars.  

\section{Discussions}
We see that radio-loud fraction is increasing with FWHM. Our main goal is to search the reason why we are getting more number of RL quasars at higher FWHM. To make the investigation robust we have done our analysis for two broad emission lines H$\beta$ (z $< 0.75$) and MgII (z $< 1.9$). High broad line (FWHM $>$ 15,000 km $s^{-1}$) sample has been chosen for our analysis.\\
We first compared their basic properties. From bolometric luminosity distributions we can see for high broad line, RL quasars have higher luminosities for both H$\beta$ and MgII. And redshift distributions of RL quasars and RQ quasars of H$\beta$ sample tells us there is no selection bias in the sample which implies RL quasars are intrinsically brighter than RQ quasars. From the redshift distributions of RL quasars and RQ quasars for MgII sample we can see in higher redshifts we are getting more RL quasars which implies RL quasars are detected more in number than RQ quasars in high redshifts. We have already predicted the RL quasars are intrinsically brighter than RQ quasars so this result is in agreement with our prediction.\\   
From BH mass distributions we can see the same thing for both the lines  i.e., for high broad line that RL quasars are more massive than RQ quasars. Now radio luminosity is commonly used as an indirect measure of the energy transported by the central engine through the jets. So we can say from our result that higher mass implies high jet production in high broad line region. Studies like \citep{liu, Gu, McLu, Ari} have shown similar results. 
\\Otical continuum luminosity and accretion rate distributions show similar trends since continuum luminosity is generated from accretion disk. We have shown a connection between accretion disk and jet production from our results. Previous studies have agreed with our prediction. Since we know that there is a connection between radio luminosity and accretion rate \citep{Bland, Falck} thus higher radio luminosity implies high accretion rate which is evident from our results also.
\bibliography{mj}%
\section{Acknowledgement}
We acknowledge support from the Department of Science and Technology through the SERB-ECR grant of Dr. Suchetana Chatterjee.
\end{document}